\documentclass[sigconf]{acmart} 
\AtBeginDocument{%
  }

\usepackage{array}
\usepackage{booktabs}
\usepackage{comment}
\usepackage{pdflscape}
\usepackage{longtable}
\usepackage{pifont}
\newcommand{\cmark}{\ding{51}}
\newcommand{\xmark}{\ding{55}}

\copyrightyear{2026}
\acmYear{2026}
\setcopyright{cc}
\setcctype{by-nc-nd}
\acmConference[ACM REP '26]{ACM Conference on Reproducibility and Replicability}{July 20--22, 2026}{Delft, Netherlands}
\acmBooktitle{ACM Conference on Reproducibility and Replicability (ACM REP '26), July 20--22, 2026, Delft, Netherlands}
\acmDOI{10.1145/3820002.3828589}
\acmISBN{979-8-4007-2778-8/2026/07}

\begin{document}

\title{Reproducibility Challenges in Computational Network Science: Evidence, Causes, and Recommendations}

\author{Akrati Saxena}
\orcid{0000-0002-7151-6309}
\affiliation{%
  \institution{Leiden Institute of Advanced Computer Science\\ Leiden University}
  \city{Leiden}
  \country{The Netherlands}
}
\email{a.saxena@liacs.leidenuniv.nl}

\renewcommand{\shortauthors}{Saxena}

\begin{abstract}
Reproducibility is essential for scientific progress, enabling validation, fair comparison, and building upon prior work. In computational network science (CNS), however, reproducibility remains limited due to missing code, inaccessible datasets, and insufficient reporting of experimental details. This paper presents a taxonomy of reproducibility in CNS, structured around artifact availability, algorithmic clarity, experimental environments, and data processing and experimental pipelines. To systematically examine these challenges, we conduct four case studies spanning diverse methodological settings: topic-based influential user detection (network science and natural language processing-based methods), influence-based community detection (pure network-based methods), influence maximization (classical, heuristic, approximation, and AI-based mixed approaches), and reinforcement learning for network analysis (learning-based methods). Across these domains, we observe a consistent lack of publicly available artifacts, particularly code and datasets, hindering verification and comparison of results. We identify key causes of this reproducibility gap, including limited incentives for sharing artifacts, data access restrictions, incomplete experimental descriptions, and a complex methodological pipeline. Finally, we outline recommendations to improve reproducibility, including mandatory artifact sharing policies, standardized benchmarks, and comprehensive reporting of experimental setups. Addressing these gaps is critical to ensure transparency, comparability, and sustained progress in computational network science.


\end{abstract}



\maketitle

\section{Introduction}

Reproducibility is a fundamental principle of scientific research \cite{ivie2018reproducibility}. The ability to reproduce experimental results allows researchers to validate findings, compare new approaches with existing methods, and build cumulative knowledge within a field. In computational disciplines, reproducibility typically requires access to the implementation of proposed methods, the datasets used for evaluation, and sufficient details about the experimental setup. Without these elements, verifying published results becomes difficult, and the reliability of reported results and improvements remains uncertain. 

In recent years, computational network science (CNS) has emerged as a key paradigm for modeling and analyzing complex systems \cite{hexmoor2014computational}. Networks are used to model large-scale real-world complex systems, where nodes denote entities and edges represent interactions, such as friendships in social networks, transactions in financial systems, or collaboration in co-authorship networks. These networks capture diverse structural properties, including directed and weighted relationships \cite{saxena2022evolving}, overlapping communities \cite{kelley2011overlapping}, hypergraphs  \cite{fischl2021hyperbench}, and multilayer structures \cite{kivela2014multilayer}. These networks are analyzed for pattern discovery, behavioral predictions, and identification of crucial structures within a system \cite{estrada2012structure}. As a result, CNS has become central to domains such as social network analysis, computational social science, finance \cite{saxena2021banking}, criminology \cite{fonhof2019characterizing, duijn2014relative, miller2018discovering}, cybersecurity \cite{pino2014network}, media \cite{aral2018social, gupta2016modeling}, epidemiology \cite{richter2024intervention, cao2022micro}, education \cite{gera2022chunk, saxena2019survey}, and public health \cite{luke2007network}.

In CNS, a wide range of network analysis algorithms have been developed to analyze network structure at multiple scales and solve network analysis tasks, such as community detection, link recommendation, centrality ranking, influence maximization, influence blocking, and anomaly detection \cite{newman2018networks}. CNS methods are often evaluated on real-world datasets and compared against prior work using standard performance metrics. In principle, such evaluations should be reproducible, allowing researchers to verify results and benchmark new approaches. In practice, however, reproducibility remains limited. A significant portion of CNS literature does not provide access to source code, datasets, or complete experimental pipelines. Even when datasets are cited, they are often not publicly available, or the preprocessing steps required to construct the network are not sufficiently described. Critical implementation details, such as hyperparameter settings and evaluation protocols, are frequently underreported. Implementation code, or even sometimes a well-written pseudocode of the algorithm, is not provided. These gaps make it difficult to replicate results, and small variations in implementation or data processing can lead to substantially different outcomes.

To systematically assess the extent of these challenges, we conduct four case studies covering diverse downstream tasks in CNS: topic-based influential user detection, influence-based community detection, influence maximization, and reinforcement learning for network analysis. These case studies span a range of methodological settings, from purely network-based approaches to hybrid methods incorporating natural language processing, machine learning, and deep learning. Our analysis reveals a consistent lack of research artifacts, particularly code and datasets, across all case studies, highlighting a broader reproducibility gap in the field. We also contacted the authors of the first case study to request code and data; only two responded, and only one shared any materials. This further highlights the limited accessibility of research artifacts, even when directly requested from the authors.

The inability to reproduce existing methods has several negative consequences for the progress of network science research. First, it limits the ability to verify the correctness of proposed algorithms and apply them in practice. Second, it prevents fair comparisons between new methods and prior work, since the original implementations are unavailable. Third, it increases the risk of reporting improvements that may arise from differences in experimental settings rather than genuine methodological advances. Additionally, the method might not be applicable for all cases.

This paper presents a position statement on reproducibility in computational network science. We propose a taxonomy of reproducibility specific to CNS, grounded in our empirical observations across multiple tasks. We further analyze the underlying factors contributing to these issues, including limited incentives for artifact sharing, dataset accessibility constraints, and incomplete reporting practices. Finally, we outline recommendations to improve reproducibility, including mandatory artifact sharing, standardized benchmarks, and more rigorous documentation of experimental pipelines. Addressing these challenges is essential to ensure transparency, enable reliable comparisons, and support sustained progress in network science research. To the best of our knowledge, this is the first work that addresses reproducibility concerns in the CNS.

The remainder of the paper is organized as follows. In Section~\ref{related_work}, we discuss the related work. Section \ref{sec_taxonomy} presents a taxonomy of reproducibility in CNS. In Section \ref{sec_case_studies}, we analyze reproducibility in CNS through four case studies covering diverse methodological settings. Sections~\ref{sec_challenges} and \ref{sec_causes} cover the major challenges observed in reproducing network science research and their underlying causes, respectively. Based on these insights, we propose a set of recommendations to improve reproducibility practices in the field, which are discussed in Section \ref{sec_recommendations}. Finally, Section \ref{sec_conclusion} concludes the paper with a summary of findings, limitations, and future directions.

\section{Related Work}\label{related_work}

Reproducibility has long been recognized as a fundamental requirement for reliable scientific research, yet concerns about the reproducibility of computational studies have increased across many disciplines \cite{baer2018responding, antelmi2023survey, tsima2023reproducibility, gundersen2018state, moody2022reproducibility}. Empirical analyses of computing research show that a substantial proportion of published results cannot be reproduced due to incomplete documentation of methods, data, or experimental procedures \cite{antelmi2023survey, gundersen2018state}. Raghupathi et al. \cite{raghupathi2022reproducibility} conducted an empirical study of computational publications and demonstrated that most papers fail to document key reproducibility variables for data, methodology, and experiments, thereby limiting an independent researcher's ability to verify the results. Similar concerns have been raised across other scientific fields, highlighting reproducibility as a widespread challenge rather than an isolated issue.

Within computational social science and network science, reproducibility is particularly challenging due to complex analytical pipelines, evolving software ecosystems, and dependencies on large-scale digital data sources. Studies have emphasized that reproducibility requires access not only to the original data and code but also to detailed documentation and stable computational environments capable of executing the same analytical workflow \cite{schoch2024computational}. The increasing reliance on heterogeneous datasets, APIs, and specialized tools further complicates reproduction efforts, especially when dependencies change or external services become unavailable.

Data availability is one of the most significant barriers in computational fields such as computational social science, social media, and network science. Many studies rely on data collected from online platforms such as social media, where restrictive APIs or platform policies govern access. Changes in API functionality or restrictions on data sharing can prevent researchers from obtaining the same datasets used in original studies, making exact reproduction difficult or impossible \cite{hutton2015toward}. These limitations highlight the structural challenges of reproducibility in network-based studies that rely on platform-generated data.

Beyond data access, several studies have highlighted broader institutional and methodological challenges. Computational social science research often involves multidisciplinary collaboration, large-scale computational infrastructure, and sensitive datasets, all of which create barriers to data sharing and transparent workflows \cite{d2025lessons}. Even when code or datasets are partially available, insufficient documentation and environment-specific dependencies frequently prevent the successful reproduction of published results~\cite{wonsil2023integrated}. 

Recent work has also explored technical and methodological solutions for improving reproducibility. Approaches such as artifact evaluation processes, provenance tracking, and standardized documentation practices have been proposed to ensure that published research artifacts, such as code, datasets, and experimental workflows, can be independently verified \cite{abrar2025reproducibility}. Similarly, reproducibility frameworks in machine learning emphasize the importance of recording training parameters, data preprocessing steps, and computational environments through provenance systems to enable the reliable re-execution of experiments. 

Researchers have also studied reproducibility badging systems \cite{frery2020badging, radha2021verifiable}, and several publication venues have introduced these badges to promote artifact sharing. Despite these initiatives, several empirical studies demonstrate that reproducibility challenges remain widespread across computational research domains. Reproduction attempts often fail due to missing datasets, incomplete implementation details, or inconsistent experimental descriptions \cite{gundersen2018state, hutton2015toward, raghupathi2022reproducibility}. These findings suggest that improving reproducibility requires both technical solutions and cultural changes within research communities, including stronger incentives to share research artifacts and to adopt open science practices.

\section{Reproducibility Taxonomy for Computational Network Science}\label{sec_taxonomy}

In the context of computational network science, reproducibility is influenced by multiple factors ranging from the availability of research artifacts to the details of experimental pipelines. To structure these challenges, we propose a taxonomy of reproducibility issues specific to CNS. The taxonomy focuses on four main dimensions: artifact availability, experimental environment, algorithmic reproducibility, and data processing pipelines.

\subsection{Artifact Availability}

Artifact availability refers to whether the essential resources required to reproduce experimental results are publicly accessible. These artifacts typically include the implementation of the proposed method, the datasets used for evaluation, and the configuration parameters required to run the experiments. In many network science studies, the absence of these artifacts significantly limits the ability of other researchers to validate reported results. 

\subsubsection{Dataset Availability}

Datasets are central to computational network science experiments, particularly for tasks such as community detection, anomaly detection, and influential user identification, which require network structures, ground-truth labels, and auxiliary information such as textual content or user profiles. Many problems, including influential user detection \cite{panchendrarajan2023topic} and fake news detection \cite{saxena2021fake}, rely on multimodal datasets that combine network data with additional contextual information. In CNS, reproducibility becomes difficult when datasets are proprietary, private, or no longer accessible. In some cases, platform restrictions limit data sharing; for example, datasets collected via X APIs (formerly Twitter) may allow redistribution only of object identifiers (e.g., user or tweet IDs), not the full data. Similarly, when datasets are manually annotated but not shared, the absence of these labels prevents verification of the reported results. 

Another major challenge arises from insufficient documentation of the data preprocessing used to construct the network and its associated metadata, making it difficult to recreate the exact dataset used in the experiments. The construction of network datasets typically involves multiple steps, such as sampling, filtering, and graph construction, which significantly affect experimental outcomes. For example, \cite{song2017temporal} samples a subgraph from the original Gowalla dataset \cite{cho2011friendship} to test their influence blocking methods, but the sampling method is not explained in the work. Similarly, some studies restrict their analysis to the largest connected component, as their methods assume connectivity, but this choice is often not explicitly stated. Additionally, different versions of the commonly used datasets further complicate reproducibility. For example, the following snapshots of DBLP datasets \cite{lange2011frequency}, \cite{yang2012defining}, and \cite{ley2002dblp} are generally referred to by the same name. If the corresponding reference or a specific snapshot is not explicitly provided, it will limit the ability to reproduce the reported results.

\subsubsection{Code Availability}

One of the most common barriers to reproducibility is the lack of a publicly available implementation. In many network science research papers, only the methods are explained, and the algorithm or pseudocode is provided, but the source code associated with the published method is not released. Even when code is shared, it may only be partially available, excluding critical components used during evaluation. For example, in network embedding-based link prediction approaches, authors may release code for generating embeddings, but the code for validating the downstream network analysis tasks, such as the link prediction model using these generated embeddings, is missing \cite{grover2016node2vec}. Another recurring issue is inadequate documentation, which makes it difficult to understand how to run or extend the provided implementation. In some cases, code repositories exist but cannot be executed due to missing dependencies, incompatible environments, or incomplete setup instructions. These limitations significantly hinder researchers' ability to reproduce and validate reported results.

\subsubsection{Model and Parameters}

Reproducing experimental results also requires detailed information about the parameters used during model training and evaluation. Missing hyperparameter settings, unspecified initialization strategies, and undocumented random seed configurations can lead to significant variations in results. In stochastic algorithms, such as centrality ranking, influence maximization, or network embedding, small differences in parameter settings may produce substantially different outcomes. In machine learning, AI, and reinforcement learning–based approaches, trained models can be shared to ease the replication of results. 

\subsection{Algorithmic Reproducibility}

Algorithmic reproducibility refers to the ability of researchers to implement a method based on its description in the paper. 

\subsubsection{Algorithm Description and Pseudocode}

A sufficiently detailed description of the algorithm should enable independent implementation, even if the code is unavailable. However, many network science publications provide incomplete or ambiguous descriptions of the algorithms or methods they propose. In several studies, the proposed method is described at a conceptual level without providing pseudocode \cite{yazdani2020new, zhang2023large}, or the provided pseudocode does not fully capture all steps of the approach. Key implementation details are often missing, such as parameter tuning procedures, hyperparameter settings, stopping criteria, and optimization strategies.

\subsubsection{Implementation Details}

Even when the algorithm description is relatively clear, differences in implementation choices can still affect results. For example, researchers may rely on different graph processing and analysis libraries, such as iGraph \cite{csardi2006igraph}, NetworkX \cite{hagberg2007exploring}, cdlib \cite{rossetti2019cdlib}, and ndlib \cite{rossetti2018ndlib}, which might use distinct internal optimizations. Some implementations may include hidden heuristics or approximations not described in the paper, making it difficult to reproduce the reported performance.

\subsection{Experimental Environment}

Even when code and datasets are available, reproducibility may still be affected by differences in the computational environment, including variations in software dependencies, hardware resources, and runtime configurations.  

\subsubsection{Software Dependency}

Many network analysis implementations rely on a variety of external libraries and frameworks. When the versions of these libraries are not specified, it becomes difficult to recreate the exact experimental setup. Over time, software dependencies may become deprecated or incompatible with newer systems. In addition, missing or incomplete installation instructions often create further obstacles for researchers attempting to reproduce experiments.

\subsubsection{Hardware Environment}

Certain algorithms in network science, particularly those involving large-scale graph processing or deep learning–based graph representation methods, may require specific hardware configurations. Differences between CPU and GPU implementations can influence runtime performance. Furthermore, some experiments require significant memory resources or distributed computing environments, and these requirements must be clearly documented in the original work. 

\subsubsection{Runtime Configuration}

Runtime configurations such as random seed initialization and parallel execution settings can also affect reproducibility. If random seeds are not fixed, stochastic algorithms may produce different results across multiple runs. Similarly, differences in parallel processing or multithreaded execution may introduce nondeterministic behavior, complicating reproducibility. However, variations in random seeds or minor changes to the hyperparameters of machine learning algorithms should not lead to qualitatively different conclusions. If they do, the proposed approach lacks robustness, and the reported findings may not be reliable.

\subsection{Data Pre-processing}

In CNS, experimental outcomes often depend heavily on the preprocessing of raw data, as briefly discussed earlier. Many methods operate on networks constructed from heterogeneous data sources, such as social media interactions, citation networks, or textual content. Without a clear description of the preprocessing pipeline, reproducing the original dataset becomes challenging.

\subsubsection{Graph Construction}

The process of constructing a network from raw data involves several design decisions. These include filtering rules used to select nodes, criteria for defining edges, and the method used to assign edge weights. In temporal networks, additional steps, such as time-window aggregation, can significantly influence the resulting graph structure. When these decisions are not clearly documented, the reproduced network may differ substantially from the one used in the original study.

\subsubsection{Feature Engineering}

Many downstream tasks in network science, particularly those involving computational social science, such as fake news detection or topic-based influential user detection, rely on additional features derived from textual or behavioral data. Feature engineering steps, such as text preprocessing, tokenization, topic modeling, and normalization, can significantly impact model performance. If these steps are not explicitly described, reproducing the experimental pipeline becomes difficult.

\subsection{Evaluation and Validation}

The evaluation and validation process is a critical component of reproducibility in computational network science, as it directly determines how the effectiveness of a proposed method is assessed and compared with existing approaches. Even when datasets and implementation code are available, differences in evaluation strategies can lead to inconsistent or non-comparable results. 

\subsubsection{Evaluation Protocol}
A clear and complete definition of the evaluation protocol is essential for reproducibility. This much includes details such as the specification of evaluation metrics, data splits (e.g., train--test or cross-validation), number of runs, and experimental procedures. In many studies, evaluation metrics are mentioned without formal definitions or detailed explanations, leading to ambiguity in how results are computed. Additionally, variations in data-splitting strategies and experimental setups can produce substantially different outcomes, making it difficult to compare results across studies. 

\subsubsection{Evaluation Implementation}
The availability of evaluation code is essential for verifying reported results. Although many papers describe evaluation methodologies, the corresponding implementations are rarely shared, forcing researchers to reimplement them, which might potentially introduce discrepancies. Providing executable evaluation scripts with clear instructions enables exact reproduction of reported performance metrics.

\subsubsection{Baselines Methods}
The effectiveness of proposed methods is evaluated by comparing them with baseline approaches. Papers should clearly describe the baseline algorithms used, including any modifications or parameter settings applied during experiments.

\subsubsection{Baseline Implementations}
Authors should provide the implementation code for all baselines alongside their method to help evaluate all methods under identical conditions. However, in many cases, baseline methods are only referenced without providing their implementation, making fair comparison difficult. 

\subsubsection{Execution Script} 
Authors should provide execution scripts for the complete evaluation pipeline. The absence of such scripts makes it difficult to regenerate published results, even when partial artifacts are available. Providing end-to-end evaluation pipelines enables reproducibility of experimental results with minimal effort.

This taxonomy highlights that reproducibility challenges in the CNS are rarely caused by a single factor. Instead, they might arise from a combination of missing artifacts, incomplete algorithm descriptions, undocumented experimental environments, and insufficiently described data pipelines. Addressing these issues requires stronger community practices around artifact sharing, documentation, and experimental transparency.

\section{Reproducibility Case-Studies}\label{sec_case_studies}

To examine reproducibility across different methodological settings in CNS, we conduct four complementary case studies. These include \textit{topic-based influential user detection}, which combines network analysis with natural language processing (NLP); \textit{influence-based community detection}, which relies purely on network topology and diffusion dynamics; \textit{influence maximization}, a classical network science problem addressed using a variety of algorithmic approaches; and \textit{reinforcement learning for network analysis}, which integrates reinforcement learning (RL) techniques for network-based tasks. In all case studies, only peer-reviewed journals and conference publications are included.

Together, these case studies span a broad spectrum of CNS methods, ranging from purely network-based approaches to hybrid pipelines that combine network science with NLP, ML (machine learning), AI (artificial intelligence), and RL-based techniques for solving network science problems. They also cover different data settings, from simple network datasets to complex multimodal data. This diversity enables a comprehensive assessment of reproducibility challenges across varying problem formulations, dataset types, methodological approaches, and experimental pipelines in CNS. 

\subsection{Topic-based Influential Users Detection}

To examine reproducibility challenges in computational network science, we conduct a case study on \textit{topic-based influential user detection}. This task aims to identify influential users in a social network for a specific topic by combining network structure with textual or topical information. Over the past decade, numerous approaches have been proposed, often integrating graph-based influence measures with topic modeling techniques \cite{panchendrarajan2023topic}. 

Despite the growing body of work, reproducing these methods in practice remains challenging. In many cases, implementations, datasets, and experimental pipelines are not publicly available, limiting the ability to validate reported results or compare newly proposed methods with existing approaches. 

\subsubsection{Research Paper Selection Strategy and Artifacts Collection}
To examine the extent of this issue, we searched ``topic-based influential user detection" on Google Scholar and collected 30 research papers. The selected papers span multiple years and venues within the fields of network science, data mining, and social network analysis; refer to Table~\ref{tab_detailed_tbiud} in Appendix~\ref{appendix_detailed_tables} for the complete list of papers.

For each paper, we manually examined the availability of key research artifacts, including a complete explanation of the method/ pipeline, citations for datasets, availability of code and datasets, initialization of all hyperparameters, specification of the hardware environment, and evaluation methods. We specifically assessed whether the implementation of the proposed methods and datasets was publicly available. Out of 30 papers, 29 did not provide any code or dataset repository in the paper or supplementary materials. To further investigate, we contacted the corresponding authors via email, requesting access to implementation code and experimental resources, explaining our objective of conducting a comparative analysis for our survey \cite{panchendrarajan2023topic}. We could obtain valid contact information for only 25 papers, and requests were sent accordingly. For the remaining papers, two had invalid email addresses, and two did not provide author contact information; these correspond to the last 4 rows in Table \ref{tab_detailed_tbiud}. Detailed results for each paper are presented in Table~\ref{tab_detailed_tbiud}, while the aggregated findings are summarized in Table~\ref{tab_summary_tbiud}.

\begin{table}[t]
\centering
\caption{Summary of research artifact availability and author response for all reviewed papers on topic-based influential user detection (Total: 30 research papers).}
\label{tab_summary_tbiud}
\begin{tabular}{|p{3.1cm} |p{4.6cm}|}
\hline
\textbf{Artifact Category} & \textbf{Availability Count and Remarks} \\ \hline

Clear description of methodology and experimental pipeline & 30/30  \\ \hline 

Dataset citation & 9/30  \\ \hline 

Dataset availability & 0/30 papers provide publicly accessible datasets \\ \hline 

Hyperparameter reporting & 6/29 papers fully report hyperparameters (1 paper not applicable) \\ \hline 

Hardware environment specification & 7/30  \\ \hline 

Evaluation protocol & 25/30 papers describe the evaluation methodology \\ \hline

Source code availability & 1/30 papers provide publicly accessible implementation \\ \hline 

Author contact attempts & 25/30 papers; valid contact information identified and emails sent \\ \hline

Author response rate & 2/25 authors responded to requests \\ \hline

Artifact sharing upon request & 1/25 responses; one author reported code unavailability due to system failure \\ \hline

\end{tabular}
\end{table}

\subsubsection{Reproducibility Analysis}

Our analysis reveals a substantial gap between methodological descriptions and reproducibility in practice. As shown in Table~\ref{tab_summary_tbiud}, all examined papers provide a high-level description of their approaches; however, the absence of essential artifacts, such as datasets, source code, and detailed experimental configurations, severely limits the ability to reproduce reported results. 

Only one paper (\cite{chen2019joint}) provides publicly accessible source code. While the repository includes documentation and execution instructions (in the readme file), as well as a sample dataset, the actual datasets used in the experiments are not shared. Furthermore, implementations of baseline methods are not provided, which restricts the ability to perform fair and consistent comparisons.

We also observe that the dataset availability is particularly limited. Only 9 out of 30 papers cite the datasets used, and none provide public access to them. Additionally, in some cases, the preprocessing steps used to construct the network or extract topic information were not fully described. In most of the papers, datasets are self-collected and manually annotated, yet neither the data nor the annotations are shared. This is a critical limitation, as many methods rely on annotated and multimodal datasets that cannot be reconstructed without access to the original resources, making independent validation infeasible. 

Incomplete reporting of experimental configurations further hinders reproducibility. Hyperparameters are fully specified in only 6 out of 29 applicable papers, and hardware or computational environment details are reported in just 7 out of 30 papers. Given the sensitivity of many methods to parameter settings, these omissions can lead to significant variations in results.

Although most papers (25/30) clearly define the evaluation protocol or methodology used to assess performance, the absence of executable pipelines and baseline implementations remains a major barrier. None of the studies provides code for baseline methods, requiring researchers to reimplement them and introducing another layer of complexity.

To assess whether reproducibility could be improved through direct author engagement, we contacted corresponding authors to request access to code and datasets. Due to the limited availability of contact information, requests were sent to 25 authors. Only 2 authors responded in a month, and only one provided code. In another case, the author indicated that the original implementation was no longer available due to a system failure. The majority of requests received no response. These observations highlight the limited accessibility of research artifacts, even when explicitly requested.

Overall, this case study demonstrates that reproducibility in topic-based influential user detection is severely constrained by missing artifacts, incomplete reporting, and limited accessibility of experimental resources. These findings emphasize the need for stronger artifact-sharing and reporting practices in this area. 

For the next three case studies, we did not contact corresponding authors, as our main aim is to analyze the availability of artifacts reported in the papers and their associated repositories. Direct outreach at this scale is time-intensive and leads to limited responses.

\subsection{Influence-Based Community Detection}

Community detection is a fundamental task in network science that aims to identify groups of nodes that interact more frequently with each other than with the rest of the network \cite{fortunato2016community, arya2022node}. Traditional community detection algorithms rely on structural properties of the network, such as edge density, modularity, or node similarity. In recent years, however, several studies have explored the use of \textit{influence propagation} or \textit{information diffusion} to identify communities in networks \cite{das2021deployment}. 

In social networks, information spreads through user interactions, forming cascades that reflect the underlying network structure. Such diffusion processes can reveal latent communities, as nodes within the same community tend to influence one another more frequently than nodes across communities \cite{das2021deployment}. 
Influence-based community detection methods use diffusion patterns to identify communities by modeling how information propagates. These methods use well-known diffusion models, such as the Independent Cascade (IC) and Linear Threshold (LT) models, and their extensions to simulate influence propagation within the network. In these models, nodes become activated when influenced by their neighbors, and communities are identified based on patterns of activation or influence spread. Existing influence-based community detection methods remain largely \textit{network-based}, mainly using structural properties and node characteristics, without incorporating ML or AI-based techniques \cite{das2021deployment}.

\subsubsection{Paper Selection Strategy and Artifact Collection}

To analyze the reproducibility of influence-based community detection methods, we first collected relevant literature using a survey by Das and Biswas \cite{das2021deployment} on influence-based community detection. We selected papers that explicitly incorporate influence propagation or information diffusion in community detection and supplemented it with more recent works by checking the citations of all these starting papers and an explicit search on Google Scholar for ``influence diffusion based community detection" and ``information propagation + community detection". 
The selection process followed three criteria:

\begin{itemize}
\item \textbf{Influence-based formulation:} The method must explicitly use influence propagation, information diffusion, or cascade dynamics as part of the community detection algorithm.
\item \textbf{Network-based methods:} Methods purely based on network topology and nodes' characteristics are considered. 
\item \textbf{Classical Network Science Approaches:} All selected methods follow classical network science approaches, without using machine learning or deep learning-based techniques.
\end{itemize}

Using this exhaustive literature search, we identified a total of \textbf{15 papers}. For each paper, we manually assessed the availability of key research artifacts required to reproduce the reported results. Detailed artifacts availability per each paper is presented in Table~\ref{tab_detailed_cd} in Appendix~\ref{appendix_detailed_tables}, and a summary is provided in Table~\ref{tab_summary_cd}.

\begin{table}[t]
\centering
\caption{Availability of research artifacts for influence-based community detection methods. The table summarizes the extent to which essential experimental artifacts are reported or made publicly available across all reviewed papers (total: 15 research papers).}
\label{tab_summary_cd}
\begin{tabular}{|p{3.1cm} |p{4.6cm}|}
\hline
\textbf{Artifact Type} & \textbf{Availability and Remarks} \\ \hline

Algorithm pseudocode & 13/15 \\ \hline

Dataset citation & 14/15 \\ \hline

Dataset availability & 1/15; One more work provides partial datasets. \\ \hline 

Hyperparameter reporting & 6/15  \\ \hline 

Hardware environment specification & 4/15; Only 5 papers provide computational environment details (e.g., CPU/GPU configuration). \\ \hline

Publicly available source code &  2/15  \\ \hline 

Baseline implementations &  1/15 \\ \hline

Execution scripts &  1/15  \\ \hline 

Evaluation protocol & 15/15; However, executable evaluation code is available only for one paper that released source code. \\ \hline

\end{tabular}
\end{table}

\subsubsection{Reproducibility Analysis}

To evaluate reproducibility, we examined the availability of key research artifacts, including algorithm pseudocode, dataset citation and sharing, hyperparameter configurations, computational environment specifications, implementation code, execution scripts, baseline implementations, and evaluation methods along with their implementations, as summarized in Table~\ref{tab_summary_cd}.

Given that the selected methods are purely network-based, algorithmic clarity is critical for reproducibility. In this context, pseudocode serves as a primary artifact for understanding and reimplementing the methods. All papers provide conceptual descriptions of their approaches, with 13 out of 15 papers including pseudocode of the proposed method. Additionally, 14 out of 15 papers cite the datasets used in their experiments, indicating that the datasets are generally identifiable from the literature. However, only \textbf{one paper} provides partial access to the datasets, whereas in most cases the datasets must be obtained from external sources or reconstructed. This places a significant burden on researchers, as both data acquisition and preprocessing pipelines need to be self-implemented and often might not be fully specified.  

Similarly, only 6 of 15 papers report the hyperparameter values used in their experiments, making exact replication of the reported results difficult. Influence dynamics-based community detection methods generally use influence propagation models such as the Independent Cascade (IC) and Linear Threshold (LT) models, which are stochastic in nature. Therefore, all hyperparameters, such as edge influence probabilities, seed selection strategies, stopping criteria, the number of simulation runs, and random seeds, can directly affect experimental outcomes and must be clearly specified to ensure accurate replication.

The specification of computational environments is similarly limited, with only 4 papers reporting hardware details such as CPU or GPU configurations. These factors can influence runtime performance and scalability, particularly for large-scale networks.  
The availability of implementation artifacts is even more limited. Only 2 out of 15 papers provide publicly available source code for their methods. However, only \textbf{one paper} provides implementations of baseline algorithms used for comparison. Execution scripts to directly execute the method are provided by both of these papers. 

Although all papers describe their evaluation methodologies at a conceptual level, executable evaluation pipelines are rarely provided. As a result, researchers attempting to reproduce the experiments must independently implement both the proposed algorithms and the evaluation framework.

Overall, this case study highlights a significant reproducibility gap in influence-based community detection. While algorithmic concepts are generally well described, the lack of accessible datasets, implementations, and detailed experimental configurations makes independent verification difficult.

\subsection{Influence Maximization}
Influence maximization (IM) is a fundamental problem in network science that aims to identify a set of seed nodes whose activation maximizes the spread of information, influence, or awareness in a network. Given a network and a diffusion model, the objective is to select a limited number of seed nodes under the given budget such that the expected number of influenced nodes is maximized. The influence propagation is modeled using the Independent Cascade, Linear Threshold, or compartment-based spreading (SI, SIR) models, or their extensions \cite{saxena2021fake_pr}, in which information spreads along edges according to probabilistic or threshold-based rules. 

Due to its combinatorial nature, influence maximization is computationally challenging and an NP-hard problem. Researchers have proposed a wide range of solution approaches, including greedy, approximation, heuristic, probabilistic, machine learning, deep learning methods, and, more recently, reinforcement learning-based methods. These approaches are evaluated using network datasets and simulation-based diffusion processes. As a result, reproducibility critically depends on the availability of datasets, algorithm implementations, diffusion model parameters, and detailed experimental configurations.

\subsubsection{Paper Selection Strategy and Artifacts Collection}

To analyze reproducibility in influence maximization, we conducted a systematic literature search using the query \textbf{``influence maximization in social networks''} on Google Scholar. From the retrieved results, we selected the first 40 papers, ensuring that they focus on influence maximization methods in social networks. These papers were used for the subsequent reproducibility analysis. 

We manually reviewed all selected papers to assess the availability of different types of research artifacts. Detailed artifact availability for each paper is presented in Table~\ref{tab_detailed_im} in Appendix \ref{appendix_detailed_tables}, while the aggregated results are summarized in Table~\ref{tab_summary_im}.

\begin{table}[t]
\centering
\caption{The table summarizes the availability of research artifacts for influence maximization in the social networks domain (total: 40 research papers). } 
\label{tab_summary_im}
\begin{tabular}{|p{3.1cm} |p{4.6cm}|}
\hline
\textbf{Artifact Type} & \textbf{Availability and Remarks} \\ \hline

Algorithm pseudocode & 33/40 \\ \hline

Dataset citation & 33/39 (one paper excluded as it does not conduct experimental evaluation). \\ \hline 

Dataset availability & 0/39; Two papers provide some datasets in their code repository. \\ \hline 

Hyperparameter reporting & 31/39  \\ \hline 

Hardware environment specification & 27/39 \\ \hline

Publicly available source code &  3/39  \\ \hline 

Baseline implementations &  1/39 \\ \hline

Execution scripts &  2/39  \\ \hline 

Evaluation protocol & 39/39  \\ \hline 

Evaluation code & 0/39. Shared code repositories do not provide the complete evaluation code. \\ \hline

\end{tabular}
\end{table}

\subsubsection{Reproducibility Analysis}

As mentioned in Table~\ref{tab_summary_im}, though most studies provide a conceptual description of the proposed algorithms, only 33 out of 40 papers present pseudocode. One paper is purely theoretical and contains no experimental evaluation. Consequently, experiment-related reproducibility criteria (e.g., evaluation metrics, hyperparameters, hardware, and datasets) are not applicable to this paper, and, therefore, research artifact availability is analyzed across all 39 applicable papers. We observe that the dataset references are commonly reported: 33 papers cite the datasets used in their experiments. However, the actual availability of datasets is extremely limited: only one paper provides datasets through its code repository. Hyperparameters are relatively well-documented, with 38 papers reporting the parameter settings used in their experiments. Only 27 papers specify the hardware or computational environment used in their experiments.

However, the availability of artifacts required for full reproducibility is very rare. Only 3 (\cite{zhang2023capacity}, \cite{razaghi2022group}, and \cite{wang2023multi}) out of 39 papers released their source code, and only one out of these three provides implementations of baseline algorithms used for comparison. Similarly, only \textbf{one paper} provides access to most of its datasets \cite{wang2023multi}, another provides only a few through its code repository, and \cite{razaghi2022group} does not provide any datasets. Execution scripts required to reproduce the experiments are available for both of these papers. Although almost all papers (39/39) describe their evaluation protocol, the actual evaluation code is rarely shared. No repository provides the complete evaluation code for reproducibility. However, one paper provides most of the evaluation code, while others that release code do not include the complete evaluation pipeline. 

Overall, despite well-documented methodological descriptions, the lack of publicly available code, datasets, and executable pipelines significantly constrains reproducibility in influence maximization research. This limits the ability to validate results, reproduce experiments, and perform fair comparisons across methods.

\subsection{RL for Network Analysis}

Reinforcement Learning (RL) has recently gained attention as a powerful framework for solving optimization and decision-making problems in network science, such as influence maximization, influence blocking, link prediction, community detection, and key node identification. Such problems can be formulated as sequential decision-making processes in which an agent interacts with a network environment to learn a policy that maximizes cumulative reward through iterative exploration \cite{gajane2022survey}. Unlike traditional approaches, which often require retraining models for each dataset, RL-based methods aim to learn transferable policies. Once trained, these policies can be applied to new networks with similar structural properties, enabling scalable solutions for large and dynamic environments~\cite{piano, saxena2025dq4fairim}.

Despite these advantages, RL-based approaches introduce additional layers of complexity that significantly impact reproducibility. The performance of RL methods depends heavily on the precise definition of the environment, including state representation, action space, and transition dynamics. The design of the reward function is critical, as it directly influences the learned policy and the agent's overall behavior. Even small variations in reward formulation, exploration strategy, or training procedures can significantly affect the experimental outcomes.

In addition, RL models involve numerous hyperparameters, such as learning rates, discount factors, exploration schedules, and the number of training episodes. Many studies also rely on stochastic processes and randomized initialization, making results sensitive to random seeds and experimental settings. In many studies, these details are either partially reported or omitted, making it difficult to replicate the training process and reproduce results. 

We include this case study to highlight reproducibility challenges in modern learning-based CNS approaches, in which increased model complexity introduces additional barriers beyond those encountered in traditional network science methods. 

\subsubsection{Paper Selection Strategy and Artifact Collection}

We conducted an extensive literature search on RL-based methods for network analysis using Google Scholar and found a total of 28 relevant research papers. To systematically evaluate reproducibility in RL-based network analysis, we assessed the availability of research artifacts as detailed in Table \ref{tab_detailed_rlforns} (in Appendix~\ref{appendix_detailed_tables}), and a summary of this analysis is presented in Table~\ref{tab_summary_rlforns}.

\begin{table}[t]
\centering
\caption{The table summarizes the availability of research artifacts for reinforcement learning (RL) based network analysis methods (total: 28 research papers).} 
\label{tab_summary_rlforns}
\begin{tabular}{|p{3.1cm} |p{4.6cm}|}
\hline
\textbf{Artifact Type} & \textbf{Availability and Remarks} \\ \hline

Algorithm pseudocode & 24/28; Two out of the remaining four papers provide only partial pseudocode descriptions, while two papers do not include pseudocode. \\ \hline

Dataset citation & 26/28; One out of the remaining two papers does not cite the dataset, and another does not fully specify all datasets used. \\ \hline

Dataset availability & 4/27; One remaining paper generates networks using the R library, and one additional paper reports that the dataset is available upon request.  \\ \hline 

Hyperparameter reporting & 4/28  \\ \hline 

Hardware environment specification & 11/28  \\ \hline 

Publicly available source code &  7/28  \\ \hline 

Baseline implementations &  0/28 \\ \hline 

Execution scripts &  7/28  \\ \hline 

Evaluation protocol & 26/28; However, executable evaluation pipelines are available only for the papers that release source code. \\ \hline

\end{tabular}
\end{table}

\subsubsection{Discussion}

As also observed in previous cases, while most studies provide high-level descriptions of their approaches, the availability of artifacts necessary for full experimental reproducibility remains limited. A majority of papers (24/28) include algorithm pseudocode, which supports understanding of the conceptual workflow. However, pseudocode alone is insufficient for reproducing RL experiments, as results depend on various hyperparameters and training configurations. Two papers provide only partial pseudocode, and two provide none, further limiting clarity.

Most studies cite the datasets used in their experiments (26/28), indicating that dataset identification is generally well documented, as also observed earlier. However, only five papers provide direct access to the datasets used in their experiments, where one work generated networks using R libraries. In addition, one paper states that the dataset can be obtained upon request. The lack of accessible datasets, combined with insufficient documentation of preprocessing steps and dataset versions, presents a significant barrier to reproducibility.

Reporting of hyperparameters is particularly limited, with only 4 of 28 papers providing full configurations. This is a critical issue for RL-based approaches, where training dynamics and final performance can be highly sensitive to hyperparameter choices such as learning rate, exploration strategy, discount factor, and training episodes. Without these details, reproducing the training process becomes extremely difficult.

Hardware environment details are reported in less than half of the studies (11/28). While hardware specifications alone do not guarantee reproducibility, they provide important context for computational performance and training time, especially for deep RL models that may require GPU acceleration.

The availability of publicly accessible source code is limited, with only 7 papers releasing their implementation. However, all of these works provide execution scripts for running the experiments. Another notable observation is that none of the papers considered provide implementations of the baseline methods used for comparison. Although 26 papers describe evaluation protocols, executable evaluation methods are available only for studies that release code. 

Overall, our analysis highlights a substantial gap between the conceptual description of RL-based methods and the practical requirements for their reproducibility. Due to the sensitivity of RL approaches to implementation and experimental settings, improving the availability of implementation code, datasets, complete configurations, baseline code, and evaluation methods is essential to ensure reliable and reproducible research in this domain.

\section{Challenges in Reproducing CNS Research}\label{sec_challenges}

Reproducing results in network science involves a combination of practical, technical, and methodological challenges. These challenges often arise from the absence of essential research artifacts, incomplete methodological descriptions, and the complexity of experimental pipelines used in network analysis tasks.

One of the most significant challenges is the lack of publicly available implementation of proposed methods. As observed in all case studies, many research papers introduce new algorithms and report performance improvements, but the corresponding source code is not released. In such cases, researchers must rely solely on textual method descriptions or pseudocode, which often omit critical design decisions and optimization details. As a result, accurately reproducing the algorithm and validating the reported results becomes difficult. 

Another challenge arises from dataset accessibility and preprocessing. Network science studies often rely on real-world datasets from social media platforms, biological systems, technological networks, and other large-scale systems. In recent years, the adoption of deep learning in CNS methods has led to the widespread use of multimodal datasets \cite{hong2022deep}. However, the datasets used in the experiments may not be publicly available due to privacy constraints, licensing restrictions, or platform policies. Even when datasets are available, the preprocessing steps required to construct the network, such as node filtering, edge weighting, node or edge labeling, temporal aggregation, and feature extraction, are often insufficiently documented. Additionally, in tasks such as topic-based influential user detection, datasets are largely manually annotated, but these labels are not shared, preventing proper validation (as shown in Table \ref{tab_summary_tbiud}). As a result, researchers attempting to reproduce experiments may construct networks that differ from those used in the original study, leading to inconsistent outcomes.

Experimental pipelines and parameter configurations further complicate reproducibility. Many network analysis methods rely on multiple hyperparameters related to model training, optimization, and evaluation. Variations in evaluation protocols, including differences in train–test splits, baseline implementations, and performance metrics, often lead to results that are not directly comparable across studies. Even minor variations in parameter values can lead to substantially different results, particularly in tasks such as influence maximization, influence blocking, community detection, and network embeddings. 

The computational environment also plays an important role in reproducibility. Differences in software libraries, framework versions, hardware configurations, and runtime settings can lead to variations in experimental outcomes. In many cases, these dependencies are not explicitly documented, making it difficult to recreate the original experimental conditions. Furthermore, some studies depend on APIs, proprietary tools, or dynamic data sources that change over time or become unavailable, making exact replication infeasible.

Reproducibility is further affected by the increasing complexity of modern CNS pipelines. Many tasks integrate multiple components, such as graph algorithms, machine learning models, and natural language processing techniques. For example, topic-based influential user detection combines network analysis with natural language processing techniques on textual data, such as topic modeling. These multi-stage pipelines involve several intermediate steps, each of which may introduce variability if the implementation details are not clearly described.

Beyond the core challenges discussed above, several additional factors further complicate reproducibility in the CNS. The presence of stochastic components, such as random initialization or sampling procedures, introduces non-determinism when random seeds are not reported. Reproducibility is also affected by dependencies and software decay, as evolving libraries and frameworks can render previously functional code unusable over time or require modifications to keep it compatible with the current version. Finally, the absence of baseline implementations forces researchers to reimplement them, introducing additional sources of variation.

\section{Causes of Reproducibility Issues}\label{sec_causes}

The challenges described above are often the result of structural and cultural factors within the research ecosystem rather than isolated oversights by individual authors. Several underlying causes contribute to the reproducibility gap observed in computational network science research.

One major factor is the lack of incentives for sharing research artifacts. Academic publication systems primarily reward novelty and performance improvements, while the effort required to release well-documented code, datasets, and experimental pipelines is often undervalued. As a result, researchers tend to prioritize rapid publication over artifact preparation and long-term maintenance.

Another contributing factor is time and resource constraints. Preparing reproducible research artifacts requires significant effort, including organizing code, documenting dependencies, cleaning datasets, and providing clear instructions for execution. Researchers working under tight deadlines and limited resources may not allocate sufficient time to prepare these materials for public release. 

Data access restrictions also play an important role. Many studies rely on datasets collected from social media platforms or proprietary sources that cannot be freely distributed due to privacy concerns or platform policies. In such cases, even well-intentioned researchers may be unable to share the exact datasets used in their experiments, limiting others' ability to reproduce the results.

In addition, insufficient reporting standards in network science publications contribute to reproducibility issues. Conference papers often have strict page limits, which can lead authors to omit detailed descriptions of experimental setups, preprocessing pipelines, or parameter configurations that might be considered well-known in the domain, assuming readers already know these details. However, without these details, reproducing the reported experiments becomes difficult.

Another cause is the increasing complexity of modern network analysis pipelines. Many contemporary methods combine graph algorithms with machine learning, deep learning, natural language processing, and reinforcement learning components. These complex pipelines involve multiple stages of processing, making it challenging to fully document every step within the constraints of a traditional research paper. Finally, a deeper understanding of these causes requires systematic investigation to help design effective strategies for improving reproducibility practices.  

\section{Recommendations for Improving Reproducibility}\label{sec_recommendations}

In computational network science, publications in conferences and journals are equally valued. Addressing reproducibility challenges in network science requires coordinated efforts from researchers, reviewers, conference organizers, journal editors, and the broader research community. Several practical steps can significantly improve the reproducibility of future studies.

First, researchers should be encouraged to share source code and complete experimental artifacts whenever possible. Public repositories hosted on platforms such as GitHub allow authors to share implementations, documentation, and executable scripts that demonstrate how to reproduce results. Providing well-structured repositories with clear instructions can greatly reduce the effort required for other researchers to replicate results. 

Second, datasets and preprocessing pipelines should be documented and shared to the extent permitted by data policies. When direct data sharing is not possible, authors should provide detailed descriptions of data collection, preprocessing, and network construction procedures to enable the reconstruction of comparable datasets. In addition, methods should be evaluated on publicly available benchmark datasets and provide these networks so that the reproducibility of the work can be tested. The network science communities still lack standardized benchmark datasets, particularly for emerging challenges such as benchmark networks with structural inequalities \cite{saxena2024fairsna}, and there is a clear need to develop and adopt such resources. The community should develop synthetic benchmark dataset models and standardized evaluation protocols that will enable consistent comparison of methods and reduce variability arising from dataset or evaluation-specific choices.

Third, authors should provide complete descriptions of experimental configurations, including hyperparameter settings, evaluation metrics, and random seed configurations. Reporting these details ensures that experiments can be replicated under the same conditions used in the original study. To enforce this, reviewers can play a more active role by explicitly assessing the completeness of reproducibility-related information and requesting missing details during the review process. 

Conference organizers and journal editors can also play a crucial role by encouraging or requiring artifact availability. The adoption of reproducibility badges and artifact evaluation tracks, as implemented by venues such as \textit{Expert Systems with Applications} and \textit{Journal of Artificial Intelligence Research} \cite{gundersen2024improving}, should be more widely adopted within the computational network science community. Asking for the submission of code, datasets, and experimental documentation as part of the publication process can significantly improve reproducibility, transparency, and reliability. 

Ultimately, improving reproducibility is essential for ensuring credible and cumulative scientific progress. By promoting open research practices, standardizing evaluation frameworks, and strengthening artifact-sharing requirements, the community can strengthen the scientific foundations of the field, ensure reliable validation of results, make fair comparisons between existing and newly proposed methods, and apply the proposed methods for practical applications. Future work should conduct systematic studies and surveys with authors, reviewers, conference organizers, and journal editors to better understand existing barriers and to design effective strategies to improve reproducibility practices.

\section{Discussion and Conclusion}\label{sec_conclusion}

In this paper, we present a structured taxonomy of reproducibility in computational network science, organizing the problem into key dimensions: artifact availability, algorithmic reproducibility, experimental environments, and data preprocessing and experimental pipelines. To ensure a comprehensive evaluation, we perform four diverse case studies that cover different methodological settings, including network-only approaches (influence-based community detection), hybrid methods including network science with NLP techniques (topic-based influential user detection), optimization problems (influence maximization), and reinforcement learning–based techniques (RL methods for network science tasks).

In all case studies, we observe a consistent pattern. The first important point to note is that public access to datasets is very limited. In some tasks, such as topic-based influential user detection, datasets are often collected and annotated by authors but not shared. In several cases, the datasets are not even cited, and in some cases, they are still not publicly accessible given their citations. The availability of implementation code is extremely limited, which makes it difficult to verify results or reuse existing methods. Even when authors share the code, the corresponding datasets are often missing, making the code insufficient for full reproduction. While execution instructions are generally provided when code is released, baseline implementations are rarely shared, forcing researchers to reimplement them to reproduce complete results. In addition, hyperparameter settings are frequently incomplete, which is particularly problematic for methods where performance is sensitive to these configurations, such as diffusion models or RL-based methods. We observe that the hardware environment used for running experiments is reported in approximately half of the papers. However, software requirements are not explicitly considered in this study, as most papers do not report details such as software versions or libraries in the manuscript; even when these are shared, they are generally specified in the code repositories. To further assess reproducibility, we conducted an additional experiment in the first case study by emailing the authors of 25 papers. The majority of emails received no response, and we received the code for only one paper. This indicates that even direct outreach does not substantially improve access to research artifacts.

We further discuss the observed challenges in reproducing computational network science studies, the underlying causes, and propose concrete recommendations to improve reproducibility practices in the field. This study has certain limitations. All papers were manually reviewed and annotated, which may introduce minor inconsistencies or unintentional errors. However, to improve the reliability of the annotations, each case study was independently verified by a second researcher with expertise in this domain. The field will benefit from automated AI-based tools for reproducibility assessment that can be integrated into publication workflows for initial checks. The observed reproducibility gap calls for systematic studies and surveys involving authors, reviewers, conference organizers, and journal editors to understand its underlying causes better. It is also important to examine differences between conferences and journals, particularly to assess the impact of page limits on reproducibility. Comparing the reproducibility of journal publications that offer reproducibility badges with those that do not can also provide valuable insights. Additionally, cross-domain comparisons can help determine whether these challenges are specific to computational network science or reflect broader issues, and can help establish more robust and reliable research standards in computational research.

\begin{acks}
The author thanks Rrubaa Panchendrarajan for coordinating correspondence with authors during our earlier survey on topic-based influential user detection \cite{panchendrarajan2023topic}, which contributed to the first case study presented in this paper. The author also thanks Fabrizio Corriera, Kirtidev Mohapatra, and Harshith Kumar Yadav for independently verifying the annotations for the second, third, and fourth case studies, respectively.
\end{acks}

\bibliographystyle{ACM-Reference-Format}
\bibliography{mybib}

\appendix

\section{Description and Analysis of Research Artifacts}\label{appendix_detailed_tables}

Here, we provide a detailed description of the research artifacts considered in our reproducibility analysis across all case studies. These artifacts are essential for understanding, implementing, and reproducing the experimental results reported in computational network science studies.

We categorize research artifacts into the following groups:

\begin{itemize}
\item \textbf{Algorithm Pseudocode:} Pseudocode is a structured, language-agnostic description of the proposed method that, if correctly followed, enables correct implementation of the algorithm. A paper is marked as yes (\cmark) if pseudocode is provided, and no otherwise (or not applicable where relevant).

\item \textbf{Dataset Citation:} It covers whether the datasets used in the experiments are clearly referenced or not in the paper. 

\item \textbf{Dataset Provided:} Whether the datasets used in the experiments are publicly shared by the authors, either through repositories or supplementary materials. If datasets are accessible through their reference, but are not directly provided by the authors, they are marked as not shared (\xmark). This is because, in many cases, the final networks are constructed through multiple preprocessing steps, and the processed datasets used in the experiments should be shared to ensure reproducibility.

\item \textbf{Hyperparameter Reporting:} The extent to which experimental parameters (e.g., learning rates, diffusion probabilities, number of iterations, random seeds) are explicitly specified in the paper, which are required to reproduce the results.

\item \textbf{Hardware Environment:} Information about the hardware configuration, including CPU and GPU specifications. 

\item \textbf{Software Environment} The software environment refers to all software components required to run the experiments, such as operating system, programming language and version, libraries and frameworks, library versions, dependencies and packages, and runtime tools.

\item \textbf{Source Code Availability:} Whether the implementation of the proposed method is publicly available.

\item \textbf{Preprocessing Scripts:} In many cases, datasets require preprocessing before applying the main algorithm (e.g., filtering networks or assigning influence probabilities). If the code is shared, this criterion is marked as yes (\cmark) if preprocessing scripts are provided with code, no (\xmark) if preprocessing scripts are not provided, and not applicable (NA) otherwise. It is also marked as NA in other cases, as without the implementation code and datasets, it is difficult to judge whether preprocessing is required.

\item \textbf{Execution Scripts:} Availability of scripts or instructions to run the experiments. 

\item \textbf{Baseline Implementations:} It indicates whether the authors provide the implementations of the baseline methods used for comparison. 

\item \textbf{Evaluation Protocol:} Description of evaluation metrics and procedures used to assess performance of the proposed method.

\item \textbf{Evaluation Code:} Availability of code used to compute evaluation metrics and reproduce reported results.
\end{itemize}

\subsection{Complete Evaluation for Case Studies}

For each case study, we provide detailed tables that report the availability of the above artifacts at the level of individual papers. We note that the correctness of the pseudocode is not evaluated, as this study focuses solely on its availability in the papers. Software requirements are not explicitly considered in our case studies, as most papers did not report details such as software versions or libraries in the manuscript; even when these are shared, they are generally specified in the code repositories.

The detailed tables corresponding to different case studies are mentioned as follows:

\begin{itemize}
\item \textbf{Topic-based Influential User Detection:} Table~\ref{tab_detailed_tbiud} presents the detailed availability of artifacts for this case study. Here, we consider whether the complete method or pipeline is described, as these approaches combine network science and NLP and involve multiple components. Pseudocode is not included as a criterion because such papers use multi-stage methods and generally do not provide it; none of the surveyed papers include it. Additionally, as only one paper provides code, which does not include baseline implementations, we therefore exclude baseline and evaluation code from the table.

\item \textbf{Influence-based Community Detection:} Table~\ref{tab_detailed_cd}
\item \textbf{Influence Maximization:} Table~\ref{tab_detailed_im}
\item \textbf{Reinforcement Learning for Network Analysis:} Table~\ref{tab_detailed_rlforns}
\end{itemize}

These tables provide a comprehensive overview of artifact availability for each selected paper and support the reproducibility analysis discussed in Section \ref{sec_case_studies} of the paper.

\begin{table*}[]
\caption{Detailed reproducibility assessment of topic-based influential user detection corresponding to the summary presented in Table~\ref{tab_summary_tbiud}. }
\label{tab_detailed_tbiud}
\begin{tabular}{|l|p{1.5cm}|p{1cm}|p{1.3cm}|p{1.4cm}|p{1.3cm}|p{1.3cm}|p{1.3cm}|p{4.7cm}|}
\hline
Ref    & Complete method/ pipeline explanation & Dataset cited & Hyper-parameters reported & Hardware environment & Evaluation protocol/ method & Code availability & Dataset provided  & Remarks                     \\ \hline
\cite{li2017hybrid}               & \cmark                                       & \xmark            & \cmark                           & \xmark                             & \cmark                        & \xmark    & \xmark                         &     Data is collected from DoubanEvent but not shared.  \\ \hline 
\cite{shi2019social}              & \cmark                                       & \cmark             & \xmark                              & \xmark                             & \cmark                        & \xmark   & \xmark                          &                   \\ \hline
\cite{dhali2020attribute}         & \cmark                                       & \cmark             & \xmark                              & \xmark                             & \cmark                        & \xmark   & \xmark                          &                                                                                                \\ \hline
\cite{farahani2017characterizing} & \cmark                                          & \cmark             & NA                             & \xmark                              & \cmark                           & \xmark       & \xmark                      &                                                                                               \\ \hline
\cite{oo2020detecting}            & \cmark                                          & \xmark             & \xmark                              & \xmark                              & \cmark                           & \xmark    & \xmark                         & The dataset is a self-collected from Twitter but not shared. \\ \hline 
\cite{qian2020detecting}          & \cmark                                          & \cmark             & \cmark                              & \cmark                              & \cmark                           & \xmark       & \xmark                      &         \\ \hline
\cite{lee2019discovering}         & \cmark                                          & \cmark             & \xmark                              & \xmark                              & \cmark                           & \xmark     & \xmark                        &                \\ \hline
\cite{pal2016discovery}           & \cmark                                          & \xmark             & \xmark                              & \xmark                              & \cmark                           & \xmark    & \xmark                         &                                                                                            \\ \hline 
\cite{ma2019finding}              & \cmark                                          & \cmark             & \xmark                              & \xmark                              & \cmark                           & \xmark    & \xmark                         &      \\ \hline
\cite{koutrouli2018finding}       & \cmark                                          & \xmark             & \xmark                              & \xmark                              & \xmark                           & \xmark     & \xmark                        &   The evaluation is done against InfluenceTracker tool and its working is not clearly explained.                                                                            \\ \hline 
\cite{zhao2019high}               & \cmark                                          & \xmark             & \xmark                              & \xmark                              & \cmark                           & \xmark       & \xmark                      &                \\ \hline
\cite{shinde2016identification}   & \cmark                                          & \xmark             & \xmark                              & \xmark                              & \xmark                           & \xmark    & \xmark                         &                                                                                                                                                          \\ \hline 
\cite{quan2019identify}           & \cmark                                          & \xmark             & \xmark                              & \cmark                              & \cmark                           & \xmark        & \xmark                     &                                                                                            \\ \hline 
\cite{yu2016identifying}          & \cmark                                          & \cmark             & \cmark                              & \xmark                              & \cmark                           & \xmark   & \xmark                          &        \\ \hline 
\cite{alp2018identifying}         & \cmark                                       & \xmark            & \xmark                              & \cmark                              & \cmark                        & \xmark     & \xmark                        & They use a self-collected Twitter dataset that is not shared.                                                                                                         \\ \hline 
\cite{eliacik2018influential}     & \cmark                                          & \xmark             & \xmark                              & \xmark                              & \cmark                           & \xmark     & \xmark                        &  Its datasets are self-collected Twitter communities rather than formally cited benchmark datasets.                                                                                                                                 \\ \hline 
\cite{zheng2020measuring}         & \cmark                                          & \xmark             & \cmark                              & \cmark                              & \cmark                           & \xmark  & \xmark                           & The datasets are manually labeled but not shared.  \\ \hline
\cite{dong2019mining}             & \cmark                                          & \xmark             & \xmark                              & \xmark                              & \xmark                           & \xmark    & \xmark                         & Use self-collected dataset, but not shared . \\ \hline 
\cite{zheng2020demand}            & \cmark                                          & \xmark             & \xmark                              & \xmark                              & \cmark                           & \xmark     & \xmark                        & use self-constructed Twitter datasets                                        \\ \hline 
\cite{santospopular}              & \cmark                                          & \xmark             & \xmark                              & \xmark                              & \cmark                           & \xmark   & \xmark                          & Self-collected dataset   \\ \hline 
\cite{vega2021probabilistic}      & \cmark                                          & \xmark             & \xmark                              & \xmark                              & \cmark                           & \xmark    & \xmark                         & Dataset sourcing, full parameter reporting across studies, and hardware details are incomplete.  \\ \hline 
\cite{subbian2016querying}        & \cmark                                          & \cmark             & \cmark                              & \xmark                              & \cmark                           & \xmark      & \xmark                       &                           \\ \hline  
\cite{zhang2017recommendation}    & \cmark                                          & \xmark             & \xmark                              & \cmark                              & \cmark                           & \xmark      & \xmark                       & datasets and full tuning details are not properly specified.                                                                               \\ \hline 
\cite{lu2019topic}                & \cmark                                          & \xmark             & \xmark                              & \xmark                              & \xmark                           & \xmark       & \xmark                      &                                                             \\ \hline  
\cite{bingol2016topic}            & \cmark                                          & \xmark             &  \xmark        & \xmark                              & \cmark                           & \xmark              & \xmark               &                                                                                           \\ \hline   
\cite{chen2019joint}    & \cmark     & \xmark    & \xmark   & \cmark  & \cmark  & \xmark  & \xmark  &   \\ \hline
\cite{su2018identifying}    & \cmark     & \xmark    & \cmark   & \xmark  & \cmark  & \cmark  & \xmark  & Authors share code and sample datasets, but do not share all datasets used in experiments and code of baselines.  \\ \hline
\cite{oro2017detecting}    & \cmark     & \cmark    & \xmark   & \xmark  & \cmark  & \xmark  & \xmark  &   \\ \hline
\cite{agness2018integrated}    & \cmark     & \xmark    & \xmark   & \xmark  & \xmark  & \xmark  & \xmark  &   \\ \hline
\cite{mittal2020social}    & \cmark     & \xmark    & \xmark   & \cmark  & \cmark  & \xmark  & \xmark  &   \\ \hline
\end{tabular}
\end{table*}

\begin{landscape}
\begin{table}[]
\caption{Detailed reproducibility assessment of influence-based community detection methods corresponding to the summary presented in Table~\ref{tab_summary_cd}. The table reports the availability of key experimental artifacts for each surveyed paper, including pseudocode (Algorithm pseudocode provided), dataset citation and accessibility, hyperparameter reporting, environment specification, source code availability, preprocessing scripts/code, baselines (Baseline implementations available), explanation of evaluation method, evaluation code, and running/execution scripts.}
\label{tab_detailed_cd}
\begin{tabular}{|l|p{1cm}|p{1cm}|p{1.3cm}|p{1.3cm}|p{1.2cm}|p{1.5cm}|p{1.3cm}|p{1.1cm}|p{1.25cm}|p{1.25cm}|p{1.2cm}|p{3cm}|}
\hline
Ref                                          & Pseudo-code           & Dataset cited         & Dataset Provided      & Hyper-parameters      & Hardware environment  & Code availability     & Pre-processing scripts & Baselines             & Evaluation protocol   & Evaluation code       & Execution script   & Remarks     \\ \hline
\cite{li2020community}          & \cmark & \cmark & \xmark & \xmark & \xmark & \cmark & NA                     & \xmark & \cmark & \xmark & \cmark & Implementation of the proposed method is shared (link: \url{https://github.com/lipzh5/MWLP}), but not the code of baselines. \\ \hline
\cite{sattari2018cascade}       & \cmark & \cmark & \xmark & \xmark & \xmark & \xmark & NA                     & \xmark & \cmark & \xmark & \xmark &  \\ \hline
\cite{yazdani2020new}           & \xmark & \cmark & \xmark & \xmark & \xmark & \xmark & NA                     & \xmark & \cmark & \xmark & \xmark &  \\ \hline
\cite{shen2010hierarchical}     & \cmark & \cmark & \xmark & \xmark & \xmark & \xmark & NA                     & \xmark & \cmark & \xmark & \xmark &  \\ \hline
\cite{sun2018overlapping}       & \cmark & \cmark & \xmark & \cmark & \cmark & \xmark & NA                     & \xmark & \cmark & \xmark & \xmark &  \\ \hline
\cite{sun2020identifying}       & \cmark & \cmark & \cmark & \xmark & \xmark & \cmark & NA                     & \cmark & \cmark & \cmark & \xmark & Code and datasets are available at \url{https://github.com/sunwww168/DCDID/blob/master/DCDID/DCDID.rar} \\ \hline
\cite{das2025modeling}          & \cmark & \cmark & \xmark & \cmark & \xmark & \xmark & NA                     & \xmark & \cmark & \xmark & \xmark &  \\ \hline
\cite{das2024leveraging}        & \cmark & \cmark & \xmark & \cmark & \xmark & \xmark & NA                     & \xmark & \cmark & \xmark & \xmark &  \\ \hline
\cite{alvari2014community}      & \cmark & \cmark & \xmark & \xmark & \xmark & \xmark & NA                     & \xmark & \cmark & \xmark & \xmark &  \\ \hline
\cite{hajibagheri2012community} & \cmark & \cmark & \xmark & \xmark & \cmark & \xmark & NA                     & \xmark & \cmark & \xmark & \xmark &  \\ \hline
\cite{sun2022influence}        & \cmark & \cmark & \xmark & \cmark & \xmark & \xmark & NA                     & \xmark & \cmark & \xmark & \xmark &  \\ \hline
\cite{prasad2025influence}      & \cmark & \cmark & \xmark & \cmark & \xmark & \xmark & NA                     & \xmark & \cmark & \xmark & \xmark &  \\ \hline
\cite{xu2023influence}          & \cmark & \cmark & \xmark & \xmark & \xmark & \xmark & NA                     & \xmark & \cmark & \xmark & \xmark &  \\ \hline
\cite{abd2022influence}         & \cmark & \xmark & \xmark & \xmark & \cmark & \xmark & NA                     & \xmark & \cmark & \xmark & \xmark &  \\ \hline
\cite{zhang2023large}           & \xmark & \cmark & \xmark & \cmark & \cmark & \xmark & NA                     & \xmark & \cmark & \xmark & \xmark &  \\ \hline
\end{tabular}
\end{table}
\end{landscape}

\onecolumn

\begin{landscape}
\begin{longtable}{|p{.5cm}|p{1cm}|p{1cm}|p{1cm}|p{1.3cm}|p{1.2cm}|p{1.3cm}|p{1.2cm}|p{1.1cm}|p{1.25cm}|p{1.25cm}|p{1.15cm}|p{3.7cm}|}
\caption{Detailed reproducibility assessment of Influence Maximization methods corresponding to the summary presented in Table~\ref{tab_summary_im}.}
\label{tab_detailed_im}\\
\hline
Ref  & Pseudo-code & Dataset cited   & Dataset Provided  & Hyper-parameters & Hardware environment &  Code availability & Pre-processing scripts & Baselines & Evaluation protocol  & Evaluation code & Execution script & Remarks \\ \hline
\endfirsthead

\hline
Ref   & Pseudo-code & Dataset cited          & Dataset Provided   & Hyper-parameters & Hardware environment &  Code availability & Pre-processing scripts & Baselines & Evaluation protocol  & Evaluation code & Execution script & Remarks \\ \hline
\endhead
\hline
\multicolumn{13}{r}{Continued on next page} \\
\hline
\endfoot
\hline
\endlastfoot
\cite{chen2009efficient}       & \cmark                          & \cmark          & \xmark                                & \cmark                           & \cmark                           & \xmark               & NA                              & \xmark                                & \cmark                        & \xmark             & \xmark            &                                                                                                                                      \\ \hline
\cite{li2020community_im}         & \cmark                          & \cmark          & \xmark                                & \xmark                           & \cmark                           & \xmark               & NA                              & \xmark                                & \cmark                        & \xmark             & \xmark            &                                                                                                                                      \\ \hline
\cite{chen2010scalable}        & \cmark                          & \cmark          & \xmark                                & \cmark                           & \cmark                            & \xmark               & NA                              & \xmark                                & \cmark                        & \xmark             & \xmark            &                                                                                                                                      \\ \hline
\cite{chen2011influence}       & \xmark                          & \cmark          & \xmark                                & \cmark                           & \cmark                           & \xmark               & NA                              & \xmark                                & \cmark                        & \xmark             & \xmark            &                                                                                                                                      \\ \hline
\cite{bucur2016influence}      & \xmark                          & \cmark          & \xmark                                & \cmark                           & \cmark                           & \xmark               & NA                              & \xmark                                & \cmark                        & \xmark             & \xmark            &                                                                                                                                      \\ \hline
\cite{jung2012irie}            & \xmark                           & \cmark          & \xmark                                & \cmark                           & \cmark                           & \xmark               & NA                              & \xmark                                & \cmark                        & \xmark             & \xmark            &                                                                                                                                      \\ \hline
\cite{guo2013personalized}     & \cmark                          & \cmark          & \xmark                                & \cmark                           & \cmark           & \xmark               & NA                              & \xmark                                & \cmark                        & \xmark             & \xmark            &                                                                                                                                      \\ \hline
\cite{li2013influence}         & \cmark                          & \cmark          & \xmark                                & \cmark                           & \xmark                            & \xmark               & NA                              & \xmark                                & \cmark                        & \xmark             & \xmark            &                                                                                                                                      \\ \hline
\cite{chen2014cim}             & \cmark                          & \cmark          & \xmark                                & \cmark                           & \cmark                           & \xmark               & NA                              & \xmark                                & \cmark                        & \xmark             & \xmark            &                                                                                                                                      \\ \hline
\cite{goyal2011celf}           & \cmark                          & \cmark          & \xmark                                & \cmark                            & \xmark                            & \xmark               & NA                              & \xmark                                & \cmark                        & \xmark             & \xmark            &                                                                                                                                      \\ \hline
\cite{chen2012efficient}       & \cmark                          & \cmark          & \xmark                                & \cmark                           & \cmark                           & \xmark               & NA                              & \xmark                                & \cmark                        & \xmark             & \xmark            &                                                                                                                                      \\ \hline
\cite{liu2012time}             & \cmark                          & \cmark          & \xmark                                & \cmark                           & \cmark                           & \xmark               & NA                              & \xmark                                & \cmark                        & \xmark             & \xmark            &                                                                                                                                      \\ \hline
\cite{jiang2011simulated}      & \cmark                          & \xmark           & \xmark                                & \cmark                           & \cmark                            & \xmark               & NA                              & \xmark                                & \cmark                        & \xmark             & \xmark            &                                                                                                                                      \\ \hline
\cite{zareie2018influence}     & \cmark                          & \cmark          & \xmark                                & \cmark                           & \cmark                           & \xmark               & NA                              & \xmark                                & \cmark                        & \xmark             & \xmark            &                                                                                                                                      \\ \hline
\cite{gong2016influence}       & \cmark                          & \cmark          & \xmark                                & \cmark                           & \cmark                           & \xmark               & NA                              & \xmark                                & \cmark                        & \xmark             & \xmark            &                                                                                                                                      \\ \hline
\cite{song2016targeted}        & \cmark                          & \cmark          & \xmark                                & \cmark                           & \cmark                           & \xmark               & NA                              & \xmark                                & \cmark                        & \xmark             & \xmark            &                                                                                                                                      \\ \hline
\cite{nguyen2013budgeted}      & \cmark                          & \cmark          & \xmark                                & \cmark                           & \cmark                           & \xmark               & NA                              & \xmark                                & \cmark                        & \xmark             & \xmark            &                                                                                                                                      \\ \hline
\cite{kundu2011new}            & \xmark                           & \cmark          & \xmark                                & \cmark                           & \xmark                            & \xmark               & NA                              & \xmark                                & \cmark                        & \xmark             & \xmark            &                                                                                                                                      \\ \hline
\cite{he2019tifim}             & \cmark                          & \cmark          & \xmark                                & \xmark                           & \cmark                           & \xmark               & NA                              & \xmark                                & \cmark                        & \xmark             & \xmark            &                                                                                                                                      \\ \hline
\cite{bozorgi2017community}    & \cmark                          & \cmark          & \xmark                                & \cmark                           & \cmark                            & \xmark               & NA                              & \xmark                                & \cmark                        & \xmark             & \xmark            &                                                                                                                                      \\ \hline
\cite{kazemzadeh2022influence} & \cmark                          & \cmark          & \xmark                                & \cmark                           & \cmark                           & \xmark               & NA                              & \xmark                                & \cmark                        & \xmark             & \xmark            &                                                                                                                                      \\ \hline
\cite{gunnecc2020least}        & \cmark                          & NA           & NA                                & NA                            & NA                            & NA               & NA                              & NA                                & NA                         & NA             & NA            & The paper is purely theoretical and contains no experimental evaluation. Therefore, it does not define any evaluation metrics.                                                                                                                                     \\ \hline
\cite{kumar2022influence}      & \cmark                          & \cmark          & \xmark                                & \xmark                           & \xmark                            & \xmark               & NA                             & \xmark                                & \cmark                        & \xmark             & \xmark            &  Learning rate, training epochs and batch size are not specified.                                                                                                                                    \\ \hline
\cite{zareie2024fuzzy}         & \cmark                          & \cmark          & \xmark                                & \cmark                           & \xmark                            & \xmark               & NA                              & \xmark                                & \cmark                        & \xmark             & \xmark            &                                                                                                                                      \\ \hline
\cite{ko2018efficient}         & \cmark                          & \cmark          & \xmark                                & \xmark                           & \xmark                            & \xmark               & NA                              & \xmark                                & \cmark                        & \xmark             & \xmark            &                                                                                                                                      \\ \hline
\cite{chen2012time}            & \cmark                          & \xmark           & \xmark                                & \cmark                           & \cmark                            & \xmark               & NA                              & \xmark                                & \cmark                        & \xmark             & \xmark            &                                                                                                                                      \\ \hline
\cite{ali2021fairness}         & \xmark                           & \cmark          & \xmark                                & \cmark                           & \xmark                            & \xmark               & NA                              & \xmark                                & \cmark       & \xmark             & \xmark            & Cited source does not provide the dataset.                                                                                           \\ \hline
\cite{csimcsek2018using}       & \xmark                           & \cmark          & \xmark                                & \cmark                           & \xmark                            & \xmark               & NA                              & \xmark                                & \cmark                        & \xmark             & \xmark            &                                                                                                                                      \\ \hline
\cite{gong2016efficient}       & \cmark                          & \cmark          & \xmark                                & \xmark                           & \cmark                           & \xmark               & NA                              & \xmark                                & \cmark                        & \xmark             & \xmark            &                                                                                                                                      \\ \hline
\cite{zhou2015location}        & \cmark                          & \xmark           & \xmark                                & \cmark                           & \xmark                            & \xmark               & NA                              & \xmark                                & \cmark                        & \xmark             & \xmark            &                                                                                                                                      \\ \hline
\cite{kianian2021efficient}    & \cmark                          & \cmark          & \xmark                                & \cmark                           & \cmark                           & \xmark               & NA                              & \xmark                                & \cmark                        & \xmark             & \xmark            &                                                                                                                                      \\ \hline
\cite{zhang2023capacity}       & \cmark                          & \cmark          & \xmark                                & \cmark                           & \cmark                           & \cmark              & NA                              & \cmark                               & \cmark                        & \xmark             & \cmark           & Shared repository only has 3 out of 6 datasets. Complete evaluation code is not provided.                                               \\ \hline
\cite{razaghi2022group}        & \cmark                          & \xmark           & \xmark                                & \xmark                           & \xmark                            & \cmark               & NA                              & \xmark                                & \cmark                        & \xmark             & \xmark            &    The code is not well documented, the baseline implementations are not provided, and the repository does not include any datasets.                                                                                                                                  \\ \hline
\cite{wang2023multi}           & \cmark                          & \cmark          & \xmark                               & \cmark                           & \cmark                           & \cmark              & NA                              & \xmark                                & \cmark                        & \xmark             & \cmark           & Code is not properly documented, All baselines' code is not given, evaluation code does not generate all results shown in the paper. The shared repository includes all datasets except the Hamsterster dataset.  \\ \hline 
\cite{kumar2023influence}      & \cmark                          & \cmark          & \xmark                                & \xmark                           & \cmark                            & \xmark               & NA                              & \xmark                                & \cmark                        & \xmark             & \xmark            &                                                                                                                                      \\ \hline
\cite{lotf2022improved}        & \cmark                          & \cmark          & \xmark                                & \cmark                           & \cmark                           & \xmark               & NA                              & \xmark                                & \cmark                        & \xmark             & \xmark            &                                                                                                                                      \\ \hline
\cite{zhuang2013influence}     & \cmark                          & \xmark           & \xmark                                & \cmark                           & \xmark                            & \xmark               & NA                              & \xmark                                & \cmark                        & \xmark             & \xmark            &                                                                                                                                      \\ \hline
\cite{atif2020fuzzy}           & \cmark                          & \xmark           & \xmark                                & \cmark                           & \cmark                           & \xmark               & NA                              & \xmark                                & \cmark                        & \xmark             & \xmark            &                                                                                                                                      \\ \hline
\cite{wang2016influence}       & \cmark                          & \cmark           & \xmark                                & \xmark                           & \xmark                            & \xmark               & NA                              & \xmark                                & \cmark                        & \xmark             & \xmark            &                                                                                                                                      \\ \hline
\cite{gursoy2018influence}	& \xmark	& \cmark	& \xmark	& \cmark	& \cmark	& \xmark	& NA	& \xmark	& \cmark	& \xmark	& \xmark	&  \\ \hline															

\end{longtable}
\end{landscape}

\twocolumn

\begin{landscape}
\begin{table}[]
\caption{Detailed reproducibility assessment of RL-based network analysis studies corresponding to the summary presented in Table~\ref{tab_summary_rlforns}.}
\label{tab_detailed_rlforns}
\begin{tabular}{|l|p{1cm}|p{1cm}|p{1cm}|p{1.3cm}|p{1.2cm}|p{1cm}|p{1.3cm}|p{1.1cm}|p{1.25cm}|p{1.25cm}|p{1.2cm}|p{3.8cm}|}
\hline
Ref                 & Pseudo-code & Dataset cited          & Dataset Provided           & Hyper-parameters & Hardware environment  & Code availability & Pre-processing scripts & Baselines & Evaluation protocol  & Evaluation code & Execution script & Remarks \\ \hline
\cite{piano}               & \cmark                           & \cmark                    & \cmark                                    & \xmark                             & \cmark                                                         & \cmark & NA                & \xmark                                 & \cmark                            & \cmark             & \cmark  &          \\ \hline
\cite{you2025influence}    & \cmark                           & \cmark                    & \xmark                                     & \xmark                             & \xmark                                                          & \xmark & NA                 & \xmark                                 & \cmark                            & \xmark              & \xmark    &         \\ \hline
\cite{ma2022influence}     & \cmark                           & \cmark                    & \xmark                                     & \xmark                            & \xmark                                                          & \xmark & NA                 & \xmark                                 & \cmark                            & \xmark              & \xmark    &         \\ \hline
\cite{chen2021contingency} & \cmark                           & \cmark                    & NA                                     & \xmark                             & \xmark                                                          & \cmark & NA                & \xmark                                 & \cmark                            & \cmark             & \cmark  &          \\ \hline
\cite{tian2020deep}        & \cmark                           & \cmark                    & \xmark                                     & \xmark                             & \cmark                                                         & \xmark & NA                 & \xmark                                 & \cmark                            & \xmark              & \xmark    &         \\ \hline
\cite{halal2025topic}      & \xmark                            & \cmark                    & \xmark                                     & \xmark                            & \cmark                                                         & \xmark & NA                 & \xmark                                 & \cmark                            & \xmark              & \xmark  &           \\ \hline
\cite{chen2023touplegdd}   & \cmark                           & \cmark                    & \xmark                                     & \xmark                             & \cmark                                                         & \cmark & NA                & \xmark                                 & \cmark                            & \cmark             & \cmark  &          \\ \hline
\cite{10705687}            & \xmark                       & \cmark                    & \xmark                                     & \xmark                             & \cmark                                                         & \xmark & NA                 & \xmark                                 & \cmark                            & \xmark              & \xmark &            \\ \hline
\cite{song2025online}      & \cmark                           & \cmark                    & \xmark                                     & \xmark                            & \xmark                                                          & \xmark & NA                 & \xmark                                 & \cmark                            & \xmark              & \xmark   &          \\ \hline
\cite{yang2024balanced}    & \cmark                           & \cmark                    & \xmark                                     & \cmark                            & \cmark                                                         & \xmark & NA                 & \xmark                                 & \cmark                            & \xmark              & \xmark    &         \\ \hline
\cite{wang2025dgn}         & \cmark                           & \cmark                    & \xmark                                     & \xmark                             & \xmark                                                          & \xmark & NA                 & \xmark                                 & \cmark                            & \xmark              & \xmark   &          \\ \hline
\cite{sun2025deep}         & \cmark                           & \cmark                    & \xmark                                     & \xmark     & \xmark                                                          & \xmark & NA                 & \xmark                                 & \cmark                            & \xmark              & \xmark  &           \\ \hline
\cite{wang2025faim}        & \cmark                           & \cmark                    & \cmark                        & \xmark                             & \cmark                                                         & \cmark & NA                & \xmark                                 & \cmark                            & \cmark             & \cmark    &        \\ \hline
\cite{fadda2024math}       & \cmark                           & \cmark                    & \cmark                                    & \cmark                            & \xmark                                                          & \cmark & NA                & \xmark                                 & \cmark                   & \cmark             & \cmark  &          \\ \hline
\cite{ou2025residual}      & \xmark                            & \xmark  & \xmark & \xmark                             & \cmark                                                         & \xmark & NA                 & \xmark                                 & \cmark                            & \xmark              & \xmark   &   Not all datasets are cited, and some are available on request.       \\ \hline
\cite{jiang2023deep}       & \cmark                           & \cmark                    & \xmark                                     & \cmark                            & \cmark                                                         & \xmark & NA                 & \xmark                                 & \cmark                            & \xmark              & \xmark  &           \\ \hline
\cite{he2022reinforcement} & \cmark                           & \cmark                    & \xmark                                     & \xmark                             & \xmark                                                          & \xmark & NA                 & \xmark                                 & \cmark                            & \xmark              & \xmark  &           \\ \hline
\cite{goindani2020social}  & \cmark                           & \cmark                    & \xmark                                     & \xmark                             & \xmark                                                          & \xmark & NA                 & \xmark                                 & \cmark                   & \xmark              & \xmark  &           \\ \hline
\cite{tang2025stop}        & \cmark                           & \cmark                    & \xmark                                     & \xmark                             & \xmark                                                          & \xmark & NA                 & \xmark                                 & \cmark                   & \xmark              & \xmark   &          \\ \hline
\cite{he2022reinforcement} & \cmark                           & \cmark                    & \xmark                                     & \xmark                             & \xmark                                                          & \xmark & NA                 & \xmark                                 & \cmark                            & \xmark              & \xmark  &           \\ \hline
\cite{he2026uncertainty}   & \cmark                           & \cmark                    & \xmark                                     & \xmark                             & \cmark                                                         & \xmark & NA                 & \xmark                                 & \cmark                            & \xmark              & \xmark   &          \\ \hline
\cite{zhong2025rumor}      & \cmark                           & \xmark                     & \xmark                                     & \xmark                            & \xmark                                                          & \xmark & NA                 & \xmark                                 & \xmark                             & \xmark              & \xmark  &  Evaluation measures are not clearly defined in the experimental setup.         \\ \hline
\cite{kundu2024rumor}      & \cmark                           & \cmark                    & \xmark                                     & \xmark                             & \cmark                                                         & \cmark & NA                & \xmark                                 & \cmark                            & \cmark             & \cmark     &       \\ \hline
\cite{hasan2025multiple}   & \cmark                           & \cmark                    & \xmark                                     & \xmark      & \xmark                                                          & \xmark & NA                 & \xmark                                 & \xmark    & \xmark              & \xmark   & Infection rate for evaluation is not defined.         \\ \hline 
\cite{wu2023gac}           & \cmark                           & \cmark                    & \cmark                                    & \cmark                            & \xmark                                                          & \cmark & NA                & \xmark                                 & \cmark                   & \cmark             & \cmark     &       \\ \hline
\cite{he2021reinforcement} & \cmark                           & \cmark                    & \xmark                                     & \xmark                             & \xmark                                                          & \xmark & NA                 & \xmark                                 & \cmark & \xmark              & \xmark     &      \\ \hline 
\cite{he2022reinforcement} & \cmark                           & \cmark                    & \xmark                                     & \xmark                             & \xmark                                                          & \xmark & NA                 & \xmark                                 & \cmark & \xmark              & \xmark  &           \\ \hline 
\cite{borkar2021opinion}   & \xmark  & \cmark   & \xmark                     & \xmark                                     & \xmark                                                          & \xmark      & NA                        & \xmark                & \cmark                                & \xmark                             & \xmark & Complete pseudocode is not provided.                      \\ \hline 
\end{tabular}
\end{table}   
\end{landscape}

\end{document}